\documentclass[twocolumn,secnumarabic,amssymb, nobibnotes, aps, prd, showpacs,amsmath]{revtex4}
\usepackage[dvipdf]{graphicx}

\usepackage{graphicx}% Include figure files
\usepackage{dcolumn}% Align table columns on decimal point
\usepackage{bm}% bold math

\begin{document}

\title{Synchronized states in chaotic systems coupled indirectly through dynamic environment}
\author{V. Resmi}
\email{v.resmi@iiserpune.ac.in}
\affiliation{Indian Institute of Science Education and Research, 
Pune - 411021, India} 
\author{G. Ambika}
\email{g.ambika@iiserpune.ac.in}
\affiliation{Indian Institute of Science Education and Research, 
Pune - 411021, India} 
\author{R. E. Amritkar}
\email{amritkar@prl.res.in}
\affiliation{Physical Research Laboratory, Ahmedabad - 380009, India}
\begin{abstract}
 We consider synchronization of chaotic systems coupled indirectly through common environment where the environment has an intrinsic dynamics of its own modulated via feedback from the systems. We find that a rich variety of synchronization behavior, such as in-phase, antiphase, complete and anti- synchronization is possible. We present an approximate stability analysis for the different synchronization behaviors. The transitions to different states of synchronous behavior are analyzed in the parameter plane of coupling strengths by numerical studies for specific cases such as R\"ossler and Lorenz systems and are characterized using various indices such as correlation, average phase difference and Lyapunov exponents.  The threshold condition obtained from numerical analysis is found to agree with that from the stability analysis.
\end{abstract}
\pacs{ 05.45.Xt }

\maketitle

\section{INTRODUCTION}

Chaotic synchronization of coupled nonlinear systems has been an area of intense research activity\cite{Pikovsky2003}. In such cases, depending on the strength and nature of coupling, the systems are capable of entering into different states of synchronization such as in-phase \cite{Ros96, Rosa98}, antiphase \cite{Liu00}, lag \cite{Ros97, Tahe99}, anticipatory \cite{Vos00}, generalized \cite{Rul95, Aba96, Koc96}, complete \cite{Fuj83, Pec90} and antisynchronization \cite{Siv01, Kim03, Zhu07}. Although all these different synchronization phenomena have been explored in biological systems also, the case of phase synchronization is more useful in explaining many complex dynamical behavior in them. Specifically, antiphase synchronization with repulsive coupling has special relevance in biological systems such as neurons and ecological webs \cite{Ran06, Sin05, Che01}.

Most of the present studies on synchronization consider mutually or unidirectionally coupled systems with or without parameter mismatch.
However, synchronization has also been achieved by a common stochastic drive in uncoupled chaotic systems \cite{Tor01, Zhou02}. In such cases, the critical strength of noise for synchronization is nearly equal to the mean size of the attractor \cite{He03}. The synchronized state thus often differs very much from the intrinsic characteristics of the individual system. Synchronization of chaotic systems by external periodic forcing where the driven system locks to the frequency of the drive has also been reported \cite{ Pik97, Pik97chaos, Park99}. So also, a weak periodic force is found to stabilize inphase synchronization in a globally coupled array of Josephson junctions\cite{Brai94}.

Further, in the context of many real world systems, synchronous behavior can occur due to interaction through a common medium. For instance, synchronization of chemical oscillations of catalyst-loaded reactants in a medium of catalyst free solution is reported where coupling is through exchange of chemicals with the surrounding medium \cite{Toth06}. So also, synchronized oscillations in genetic oscillators occur due to coupling by diffusion of chemicals between cells and extracellular medium \cite{Kuz05, Wang05}. Global oscillations of concentration of neurotransmitter released by each cell can stimulate collective rhythms in a population of circardian oscillators \cite{Gon05}. Moreover, in an ensemble of cold atoms interacting with a coherent electromagnetic field, by controlling field cavity detuning, synchronized behavior with self-pulsating periodic and chaotic oscillations are found to occur \cite{Jav08}. In all these cases, the coupling function has a dynamics modulated by the system dynamics.

In general, such cases occur due to the common medium interacting with the dynamical systems. One refers to such a scheme as a coupling via a common environment.
The dynamics of $n$ systems $x_i, \; i=1,...,n$ coupled through an environment $y$ is then given by
\begin{subequations}
\label{base-model}
\begin{eqnarray}
 \dot{x}_{i} & = & f(x_i, y)  \\
 \dot{y} & = & g(y) + h({x_1,x_2,...,x_n})
\end{eqnarray} 
\end{subequations}
where $x_i$ and $y$ have dimensions $m_x$ and $m_y$ respectively. 
Such an indirect coupling has been reported in the context of periodic oscillators by Katriel \cite{Kat08}. Under suitable conditions the periodic
oscillators can synchronize.

In this paper, we consider two chaotic systems coupled through a common dynamic environment as in Eq.~(\ref{base-model}). We show that this coupling can lead to a rich variety of synchronous behavior such as antiphase, in-phase, identical, antisynchronization etc. This mechanism has the interesting feature that the common environment while capable of synchronizing the systems, does not cause major changes in their dynamics. In the synchronized state, the systems retain more or less the same phase space structure of the uncoupled system. We present an approximate stability analysis for the stability of the different synchronized states. We report detailed exploratory numerical studies for two standard systems, R\"ossler and Lorenz, and demonstrate the rich synchronization behavior. The transition to different stages of synchronization is studied by computing average phase differences, correlations, and Lyapunov exponents. From the numerical studies, we verify the relation between the critical parameters for the transition to different synchronization states obtained from the stability analysis.

\section{Environmental coupling}
We consider two chaotic systems coupled to a common environment through a linear coupling
\begin{subequations}
\label{model}
\begin{eqnarray} \label{eq:model1} 
  \dot{x}_{1} & = & f(x_1) + \epsilon_1 \gamma \beta_1 y \\ 
\label{eq:model2} 
  \dot{x}_{2} & = & f(x_2) + \epsilon_1 \gamma \beta_2 y \\
\label{eq:model3} 
  \dot{y} & = & -\kappa y - \frac{\epsilon_2}{2} \gamma^T ( \beta_1 x_1 + \beta_2 x_2)
\end{eqnarray}
\end{subequations}
The intrinsic dynamics of the environment is decaying with $\kappa$ as the damping parameter and therefore, without feedback from the systems, it is incapable of sustaining itself for extended periods of time. Here, $\epsilon_1$ is the strength of feedback to the system and $\epsilon_2$ that to the environment. For simplicity, we take $y$ to be one dimensional environment. Then, $\gamma$ is a column matrix ($m_x \times 1$), with elements zero or one, and it decides the components of $x_i$ that take part in the coupling. 

\begin{figure}
\includegraphics[width=0.95\columnwidth]{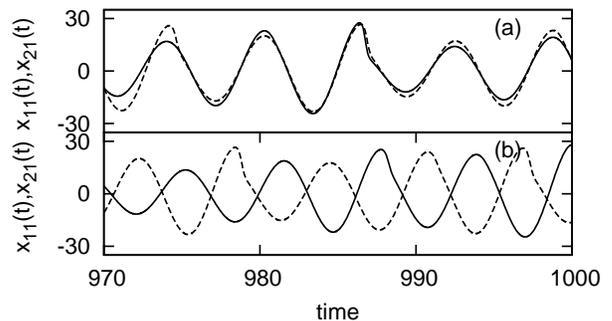}
\caption{ \label{rosts}Time series of the first variable $x_{i1}$ of two environmentally coupled chaotic R\"ossler systems showing synchronization phenomena (a) In-phase synchronization $(\epsilon_1 =  \epsilon_2 = 0.2, \; \beta_1=-\beta_2=1)$ (b) antiphase synchronizatin  $(\epsilon_1 = \epsilon_2 = 0.2, \; \beta_1=\beta_2=1)$. In both (a) and (b), we consider coupling only though one variable of the system, that is $\gamma_1 = 1$, $\gamma_i = 0$ for $i \neq 1$. R\"ossler parameters are $a = b = 0.1, \, c=18$, i.e. we have chaotic attractor and the damping parameter, $\kappa=1$.}
\end{figure}

The nature of feed back from and to the environment is adjusted by prescribing values for $\beta_1$ and $\beta_2$. When both $\beta_1$ and $\beta_2$ are of the same sign, i.e. $ (\beta_1, \beta_2) = (1, 1)$, the coupling is repulsive and can drive the systems to antiphase synchronization. When  $\beta_1$ and $\beta_2$ are of different signs, i.e. $ (\beta_1, \beta_2) = (1, -1)$, the coupling is of difference type leading to in-phase synchronization. We illustrate this behavior for the case of two chaotic R\"ossler systems coupled through environment as given by the equations
\begin{eqnarray}
\dot{x}_{i1} & = & -x_{i2} - x_{i3} + \epsilon_1 \beta_i y 
\nonumber \\
\dot{x}_{i2} & = & x_{i1} + a x_{i2} 
\nonumber \\
\dot{x}_{i3} & = & b + x_{i3} ( x_{i1} -c ) 
\nonumber \\
\dot{y} & = & -\kappa y - \frac{\epsilon_2}{2} \sum_{i=1,2}{ \beta_i x_{i1}}
 \label{eq:rossler}
\end{eqnarray}
The time series of the coupled R\"ossler systems for the in-phase synchronized and antiphase synchronized cases is shown in Fig.~\ref{rosts}a and \ref{rosts}b. 

\begin{figure}
\includegraphics[width=0.95\columnwidth]{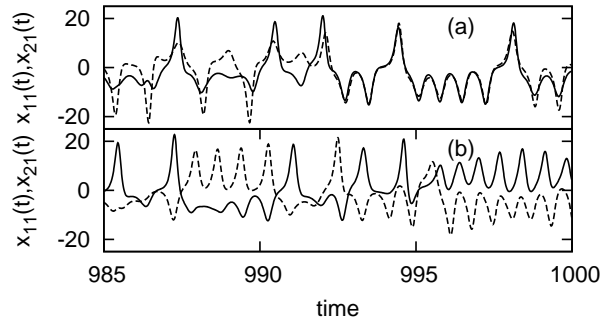}
\caption{ \label{lorts} Time series of the first variable $x_{i1}$ of two environmentally coupled chaotic Lorenz systems showing synchronization phenomena (a) In-phase synchronization $(\epsilon_1 =  \epsilon_2 = 9.0, \; \beta_1=-\beta_2=1)$  (b) antiphase synchronization  $(\epsilon_1 = \epsilon_2 = 8.0, \; \beta_1=\beta_2=1)$. Here, Lorenz parameters are $(\sigma = 10, r = 28, b = 8/3)$.}
\end{figure}

In the same way, two Lorenz systems are coupled through environment as
\begin{eqnarray}
\dot{x}_{i1} & = & \sigma (x_{i2}-x_{i1}) + \epsilon_1 \beta_i y 
\nonumber \\
\dot{x}_{i2} & = & (r-x_{i3})x_{i1} -x_{i2} 
\nonumber \\
\dot{x}_{i3} & = & x_{i1} x_{i2} - b x_{i3} 
\nonumber \\
\dot{y} & = & -\kappa y - \frac{\epsilon_2}{2} \sum_{i=1,2}{ \beta_i x_{i1}}
 \label{eq:lorenz}
\end{eqnarray}
The in-phase and antiphase synchronized states of the coupled Lorenz systems are shown in Figs. \ref{lorts}a and \ref{lorts}b.

This type of coupling is very relevant in the case of biological systems such as neurons where they interact through chemicals in the surrounding medium. We consider the case of two Hindmarsh-Rose model of neurons coupled through a common medium given by the following equations
\begin{eqnarray}
\dot{x}_{1,2} & = &  y_{1,2} + a x^2_{1,2} - x^3_{1,2} - z_{1,2} + I + \epsilon_1 \beta_{1,2} w
\nonumber \\
\dot{y}_{1,2} & = & 1 - b x^2_{1,2} - y_{1,2}
\nonumber \\
  \dot{z}_{1,2} & = & -r z_{1,2} + s r ( x_{1,2} + c )
\nonumber \\
  \dot{w}& = & - \kappa  w - \frac{\epsilon_2}{2} \sum_{i=1,2}{ \beta_i x_{i1}}
\label{eq:hrneuron}
\end{eqnarray}
 
\begin{figure}
\includegraphics[width=0.95\columnwidth]{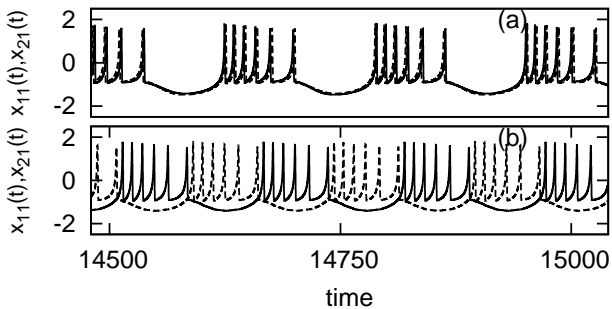}
\caption{ \label{hrts} Time series of the first variable $x_{i1}$ of two environmentally coupled Hindmarsh Rose neurons showing synchronization of bursts (a) In-phase synchronization $(\epsilon_1 =  \epsilon_2 = 0.4, \; \beta_1=-\beta_2=1)$  (b) antiphase synchronization  $(\epsilon_1 = \epsilon_2 = 0.4, \; \beta_1=\beta_2=1)$. Here, the parameters of the individual neuron are $a=3$,$b=5$,$r=0.005$,$s=4$,$c=1.6$,$I=3.05$ such that the individual neurons are chaotic. The synchronized state obtained is periodic in this case.}
\end{figure}

The in-phase and antiphase synchronized states of bursts for coupled HR neurons are shown in Fig.~\ref{hrts}a and \ref{hrts}b.

\section{Linear stability analysis}
We analyze the stability of the synchronized state of two systems coupled via the scheme of Eq.~(\ref{model}). If $\xi_1$, $\xi_2$, and $z$ represent the deviations from the synchronized state, their dynamics is governed by the linearized equations obtained from Eqs. (\ref{model}). That is
\begin{subequations}
\label{dyn-xi}
\begin{eqnarray} \label{dyn-xi1}
  \dot{\xi_1} & = & f'(x_1)\xi_1 + \epsilon_1 \gamma \beta_1 z
\\ \label{dyn-xi2}
  \dot{\xi_2} & = & f'(x_2)\xi_2 + \epsilon_1 \gamma \beta_2 z
\\ \label{dyn-z}
  \dot{z} & = & -\kappa z - \frac{\epsilon_2}{2} \gamma^T (\beta_1 \xi_1+ \beta_2 \xi_2)
\end{eqnarray}
\end{subequations}
In general, it is difficult to analyze the stability of the synchronized state from Eqs.~(\ref{dyn-xi}). For the special case of the perfectly synchronized state, i.e. $x_1=x_2$, Eqs.~(\ref{dyn-xi}) can be simplified by defining
\begin{equation}
\label{def-xi0}
\xi_0 = \beta_1 \xi_1 + \beta_2 \xi_2.
\end{equation}
Then Eqs.~(\ref{dyn-xi}) can be written as
\begin{subequations}
\label{dyn-red}
\begin{eqnarray} \label{dyn-red-xi0}
  \dot{\xi_0} & = & f'(x_1)\xi_0 + \epsilon_1 (\beta_1^2 + \beta_2^2) \gamma z
\\ \label{dyn-red-z}
  \dot{z} & = & -\kappa z - \frac{\epsilon_2}{2} \gamma^T \xi_0
\end{eqnarray}
\end{subequations}
The synchronized state corresponding to the fixed point $(0, 0)$ of
Eqs.~(\ref{dyn-red}) will be stable if all the Lyapunov exponents obtained from Eqs.~(\ref{dyn-red}) are negative.

Considerable progress can be made if we assume that the time average values of $f'(x_1)$ and $f'(x_2)$ are approximately the same and can be replaced by an effective constant value $\lambda$. In this approximation we treat $\xi_1$ and $\xi_2$ as scalars. This type of approximation was used in Ref.~\cite{amb09} and it was noted that it describes the overall
features of the phase diagram reasonably well. Thus, using $\xi_0$ defined by Eq.~(\ref{def-xi0}), Eqs.~(\ref{dyn-xi})
can be written as
\begin{subequations}
\label{dyn-approx-xi}
\begin{eqnarray} \label{dyn-approx-xi1}
  \dot{\xi_0} & = & \lambda \xi_0 + 2 \epsilon_1 z
\\ \label{dyn-approx-z}
  \dot{z} & = & -\kappa z - \frac{\epsilon_2}{2} \xi_0
\end{eqnarray}
\end{subequations}
where we choose $\beta_1^2 + \beta_2^2 = 2$.
Eliminating $z$ from Eqs.~(\ref{dyn-approx-xi1}) and~(\ref{dyn-approx-z}), we get an equation for $\xi_0$ as
\begin{equation} \label{ddot-xi0}
  \ddot{\xi_0} = ( \lambda - \kappa ) \dot{\xi_0} + ( \kappa \lambda - \epsilon_1 \epsilon_2) \xi_0
\end{equation}
Assuming a solution of the form
\begin{displaymath}
  \xi_0 = A e^{ m t }
\end{displaymath}
we get
\begin{equation}
\label{eq:m}
  m = \frac{( \lambda - \kappa ) \pm \sqrt{ ( \lambda - \kappa)^2 - 
4 (\epsilon_1 \epsilon_2 - \lambda \kappa)}}{2}
\end{equation}
The synchronized state, defined by $\xi_0 = \beta_1 \xi_1 + \beta_2 \xi_2 = 0$, is stable if Re$[m]$ is negative for both the solutions. This gives the following criteria for the stability of the synchronized state. \\
1. If $( \lambda - \kappa)^2 < 4 (\epsilon_1 \epsilon_2 - \lambda \kappa) $, $m$ is complex and the condition of stability is $\kappa > \lambda $.
\\
2. If $( \lambda - \kappa)^2 > 4 (\epsilon_1 \epsilon_2 - \lambda \kappa) $, $m$ is real and the stability condition becomes $\epsilon_1 \epsilon_2 > \lambda \kappa$ and $\kappa > \lambda $. 

In the first case above, the synchronized state is possible if we have an environment which has a sufficiently fast decay to compensate for the
divergence of the system due to $\lambda$. In the second case, an additional condition must be satisfied. Here, the transition to stable synchronization is given by the threshold values of parameters satisfying the condition
\begin{equation}\label{critical-ep}
  \epsilon_{2c} = \frac{\lambda \kappa }{ \epsilon_{1c}}
\end{equation}

We now consider the properties of the synchronized state defined by $\xi_0 = \beta_1 \xi_1 + \beta_2 \xi_2 = 0$, i.e. $\beta_1 x_1 + \beta_2 x_2 = \textrm{const.}$. Numerical simulations show that the constant is zero.
Thus, for $\beta_1=\beta_2=1$ we get $x_1=-x_2$, i.e. an antiphase synchronization while for $\beta_1=-\beta_2=1$ we get $x_1=x_2$, i.e. an in-phase synchronization.

\section{ Numerical analysis}
The scheme of coupling through the environment given in Eqs.~(\ref{model}) is applied to standard R\"ossler and Lorenz systems.  We study the two cases, $\beta_1 = +1$ and $\beta_2 = -1$ where in-phase synchronization is possible and $\beta_1 = \beta_2 = +1$ where antiphase synchronization is possible.

When $ \beta_1 = +1$ and $\beta_2 = -1$, we observe in-phase synchronization in both R\"ossler and Lorenz systems (Figs.~\ref{rosts}(a) and \ref{lorts}(a)). As the coupling strength is increased, systems go to a state of complete synchronization. When $\beta_1 = \beta_2 = +1$ the synchronized states are out of phase with each other giving antiphase synchronization for both R\"ossler and Lorenz (Figs.~\ref{rosts}(b) and \ref{lorts}(b)). As the strength of feedback is increased in the case of R\"ossler systems, control of chaos is observed and the systems become periodic, but the two coupled systems are still in antiphase synchronization. In the case of Lorenz systems as the coupling strength is increased, the systems become anti-synchronized where $x_1 = -x_2$, $y_1 = -y_2$ and $z_1 = z_2$.

\subsection{In-phase (or Antiphase) synchronization}
The transitions to in-phase ( or antiphase ) synchronization can be studied numerically using the average phase difference between the two systems. For this, we need to define phases of individual systems. In the case of R\"ossler systems, as the trajectory has a rotation around a fixed point in the $x-y$ plane, the phase $\phi (t)$ of the R\"{o}ssler system can be defined \cite{Boc02} as the angle
\begin{equation}
\label{eq:phaseros}
\phi (t) =\tan^{-1}(y(t)/x(t))
\end{equation}
The phase $\phi(t)$ and the phase difference $\psi(t)$ between the two R\"ossler systems coupled through environment are calculated using Eq.~(\ref{eq:phaseros}) for increasing strengths of feedback for identical feedback strengths $\epsilon_1 = \epsilon_2 $. The mean phase difference over many cycles $<\psi(t)>$, is nearly $0$ for the in-phase synchronization and $\pi$ for the antiphase synchronization.

Since the Lorenz system does not have such a proper rotation around any fixed point, the phase cannot be defined by Eq.~(\ref{eq:phaseros}). The phase of Lorenz system is calculated using the modified variables \cite{Pik97} as
\begin{equation}
\label{eq:phaselor}
\phi(t) = \tan^{-1}(\bar{z} / \bar{u})
\end{equation}
where $\bar{u} = u - u_p$, $\bar{z} = z - z_p$ and $u_p = \sqrt{ 2 \beta (\rho-1)}$, $z_p = \rho-1$ and $u = \sqrt{ x^2 + y^2 }$. The dynamics in $(u,z)$ looks like a rotation around some center point $(u_p, z_p)$. 
The phase $\phi(t)$ of the individual Lorenz systems are calculated using Eq.~(\ref{eq:phaselor}). The phases show confinement due to coupling indicating in-phase ( or antiphase ) synchronization. It is evident that since we neglect the sign of $x$ and $y$ in the calculation of $u$,  phase defined as in Eq.~(\ref{eq:phaselor}) can not distinguish between in-phase and antiphase cases. In this context, the similarity function $S$ \cite{Ros97} and a modified similarity function $S'$ \cite{Sen09} serves as a useful index for identifying the in-phase or antiphase synchronization.

The similarity function $S$ is defined for a delay time $\tau$
\begin{equation}
S^2(\tau) = \frac{<[x_2(t+\tau)-x_1(t)]^2>}{[<x_1^2 (t)><x_2^2 (t)>]^{1/2}}
\label{eq:simil}
\end{equation}
and the modified similarity function $S'$ is defined as
\begin{equation}
S^2(\tau) = \frac{<[x_2(t+\tau)+x_1(t)]^2>}{[<x_1^2 (t)><x_2^2 (t)>]^{1/2}}
\label{eq:modifiedsimil}
\end{equation}
For $\beta_1 = 1, \beta_2 = -1$, at $\tau = 0$, $S=0$ corresponding to the complete synchronization and $S$ is finite for the in-phase synchronization. Similarly, for $\beta_1 = \beta_2 = 1$, at $\tau = 0$, $S'$ is $0$ indicating the antisynchronization and $S'$ is finite for the antiphase synchronization.
\begin{figure}
\includegraphics[width=0.95\columnwidth]{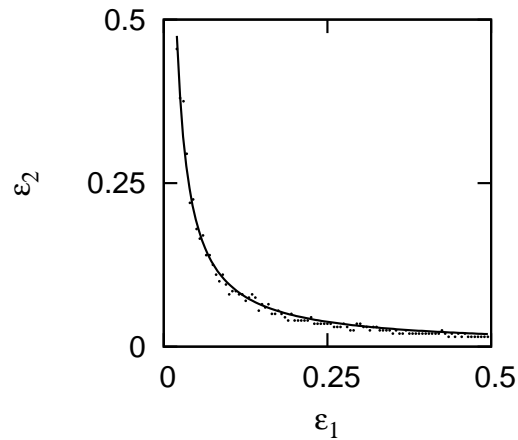}
\caption{\label{nstops_ros} Transition from regions of no synchronization to antiphase synchronization is shown in the parameter plane $\epsilon_1 - \epsilon_2$  for the coupled R\"ossler systems. The points are obtained numerically when the phase difference becomes approximately $\pi$. Solid curve corresponds to the stability condition Eq.~(\ref{critical-ep}), i.e. $\epsilon_{2c} \propto 1/\epsilon_{1c}$.}
\end{figure}
For the coupled R\"ossler systems the average phase difference is calculated for the full parameter plane $(\epsilon_1, \epsilon_2)$ in the range $(0,0.5)$ and the points where the value becomes approximately $\pi$ is plotted in Fig.~\ref{nstops_ros}. These therefore correspond to the threshold values for onset of stability of antiphase synchronizaton. The full line corresponds to the curve  plotted using the threshold condition from our stability theory in Eq.~(\ref{critical-ep}). The agreement is quite good with a $\lambda = 0.009$ and the relation $\epsilon_{2c} \propto 1/\epsilon_{1c}$ is clearly seen. Similar transition curves are also observed for transition to in-phase synchronization in the case of $\beta_1 = +1$ and $\beta_2 = -1$ and also for Lorenz systems and they agree with the relation $\epsilon_{2c} \propto 1/\epsilon_{1c}$ obtained from the stability theory.

\begin{figure}
\includegraphics[width=0.95\columnwidth]{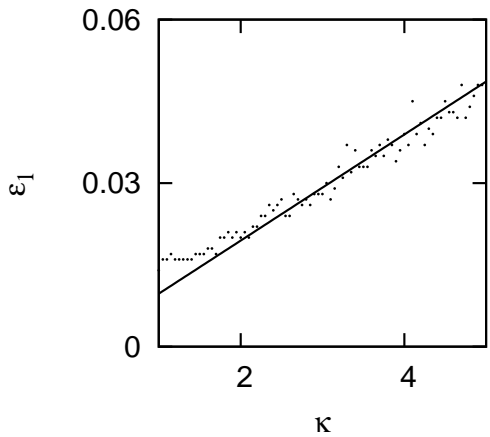}
\caption{\label{nstops_kap} Transition from unsynchronized to antiphase synchronized regions is shown in the parameter plane ($\kappa, \epsilon_1$) for coupled R\"ossler systems. Points are obtained from numerical simulation with $\epsilon_2 = 0.6$ and the solid curve is a linear fit corresponding to the stability condition Eq.~(\ref{critical-ep}) with the effective $\lambda = 0.009$.}
\end{figure}

As seen in Eq.~(\ref{critical-ep}), we also have the relations $\epsilon_{2c} \propto \kappa$ and $\epsilon_{1c} \propto \kappa$. Figure.~\ref{nstops_kap} shows the phase plot for the transition from unsynchronized to antiphase synchronized state in the $\epsilon_1 - \kappa$ plane. A linear relation is clearly seen and the solid line is drawn with the effective $\lambda = 0.009$, thus validating the transition criterion of Eq.~(\ref{critical-ep}) obtained from the stability theory.

\subsection{Lyapunov exponents}
The transitions to all the different types of synchronization described above can be tracked by calculating the Lyapunov exponents. Since the coupling here is indirect and through an environment, instead of calculating transverse Lyapunov exponents about the synchronized state, we calculate all the Lyapunov exponents of the coupled system. The variation of these Lyapunov exponents with coupling strength helps to identify the onset of in-phase ( or antiphase ) and complete ( or anti) synchronization. The two chaotic systems and the environment together form a seven-dimensional system.
The changes in the largest four Lyapunov exponents are used to identify transitions to different synchronization regimes. First crossing from zero to negative of the fourth Lyapunov exponent indicates the onset of in-phase ( or antiphase ) synchronization and the crossing of the second largest Lyapunov exponent indicates the onset of complete ( or anti ) synchronization \cite{Pikovsky2003}.
\begin{figure}
\includegraphics[width=0.95\columnwidth]{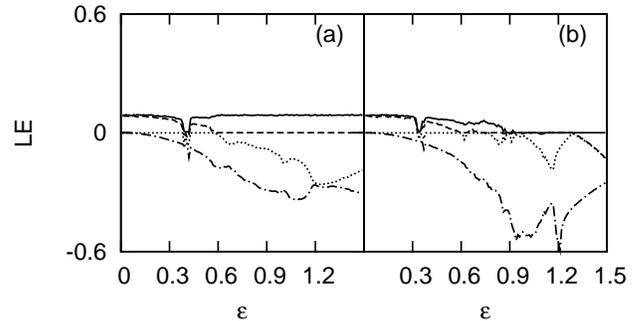}
\caption{\label{lyapunov_ros}Four largest Lyapunov exponents are shown as a function of the feed back strength $\epsilon$ for the two coupled R\"ossler systems coupled through dynamic environment with $\epsilon_1 = \epsilon_2 = \epsilon$. (a) $\beta_1 = +1, \beta_2 = -1$; the first crossing of $0$ at $\epsilon = 0.12$ (fourth largest LE) indicates the transition to in-phase synchronization, while the second zero crossing at $\epsilon = 0.59$ (second largest LE) indicates the transition to complete synchronization (b) $\beta_1 = \beta_2 = +1$; the first crossing of $0$ at $\epsilon = 0.12$ indicates antiphase synchronization and the region where all Lyapunov exponents are less than or equal to zero indicates the antiphase synchronized periodic states. (Lyapunov exponents are calculated by considering variational equations using Wolf algorithm \cite{Wolf85}.)}
\end{figure}
The largest four Lyapunov exponents for coupled R\"ossler systems are shown in  Fig.~\ref{lyapunov_ros} for various strengths of feedback. For the case $\beta_1 = +1$ and $\beta_2 = -1$, the zero crossing of the fourth largest Lyapunov exponent in Fig.~\ref{lyapunov_ros}(a) corresponds to the onset of in-phase synchronization, and the zero crossing of the second largest Lyapunov exponent corresponds to the onset of complete synchronization. Here, the narrow window where all Lyapunov exponents are less than or equal to zero corresponds to synchronized periodic states in R\"ossler systems as verified from the time series. In the case of antiphase synchronization similar results are seen (Fig.~\ref{lyapunov_ros}(b)). The region where all Lyapunov expoents are less than or equal to zero in Fig.~\ref{lyapunov_ros}(b) corresponds to the antiphase synchronization in the periodic state.
\begin{figure}
\includegraphics[width= 0.95\columnwidth]{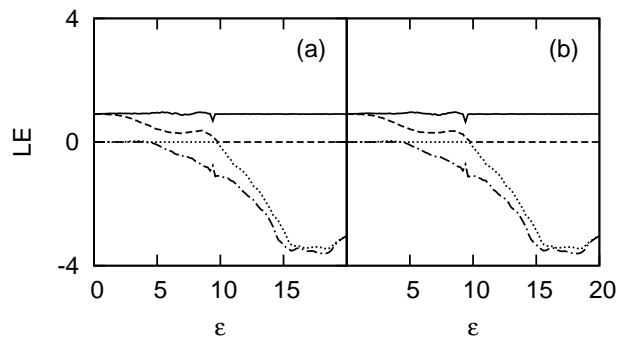}
\caption{\label{Lorlyap} Four largest Lyapunov exponents are shown as a function of the feedback strengths $\epsilon$ of two coupled Lorenz systems with $\epsilon_1 = \epsilon_2 = \epsilon$. (a) $\beta_1 = +1, \beta_2 = -1$; the first zero crossing of Lyapunov exponent at $\epsilon = 4.2 $ indicates in-phase synchronization, the second zero crossing at $\epsilon = 9.8$ indicates complete synchronization (b) $\beta_1 =  \beta_2 = +1$; the first zero crossing at $\epsilon = 4.4$ indicates antiphase synchronization, the second zero crossing at $\epsilon = 9.8$ indicates antisynchronization.}
\end{figure}
The results of a similar analysis for Lorenz are shown in Fig.~\ref{Lorlyap}. In Fig.~\ref{Lorlyap}(a), the case $\beta_1 = +1$ and $\beta_2 = -1$ is shown where the zero crossing of the fourth largest Lyapunov exponent corresponds to the onset of in-phase synchronization, and the zero crossing of the second largest Lyapunov exponent corresponds to the onset of complete synchronization. In Fig.~\ref{Lorlyap}(b), the case $\beta_1 = \beta_2 = 1$ is shown where the zero crossing of the fourth largest Lyapunov exponent corresponds to the onset of antiphase synchronization, and the zero crossing of the second largest Lyapunov exponent corresponds to the onset of antisynchronization.

\subsection{ Phase diagram }
\begin{center}
\begin{figure}
\includegraphics[width=0.95\columnwidth]{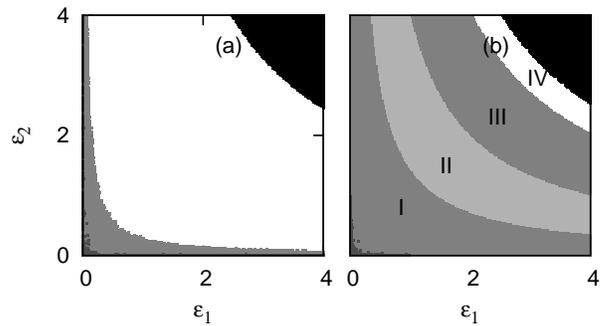}
\caption{\label{corr_ros} Regions of different states of synchronization marked out in the parameter plane  $(\epsilon_1 - \epsilon_2)$ for the coupled R\"ossler systems. The different phase space regions are obtained by using the asymptotic correlation values, average phase differences and Lyapunov exponents. (a) $\beta_1 = +1,\beta_2 = -1$. White region corresponds to $|C| \sim 0.99 $ indicating synchronized regions; light gray region is in-phase synchronized region. (b) $\beta_1 = \beta_2 = +1$ region I corresponds to antiphase synchronized chaotic states, regions II - IV corresponds to different states of antiphase synchronized periodic states (see text). In both cases, the dark gray region corresponds to the unsynchronized states and the black regions in the upper right corner are the unstable states.}
\end{figure}
\end{center}

\begin{center}
\begin{figure}
\includegraphics[width=0.95\columnwidth]{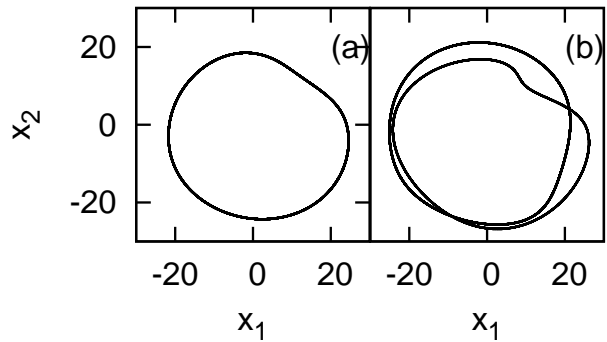}
\caption{\label{tworos} The $x-y$ phase plane of antiphase synchronized periodic states in regions II and III of Fig.~\ref{corr_ros}b. (a) $\epsilon_1= \epsilon_2 = 1.5$ Both systems are in state $A$ (b) $\epsilon_1= \epsilon_2 = 2.5$ Systems are in different states $A$ and $B$.}
\end{figure}
\end{center}

\begin{center}
\begin{figure}
\includegraphics[width=0.95\columnwidth]{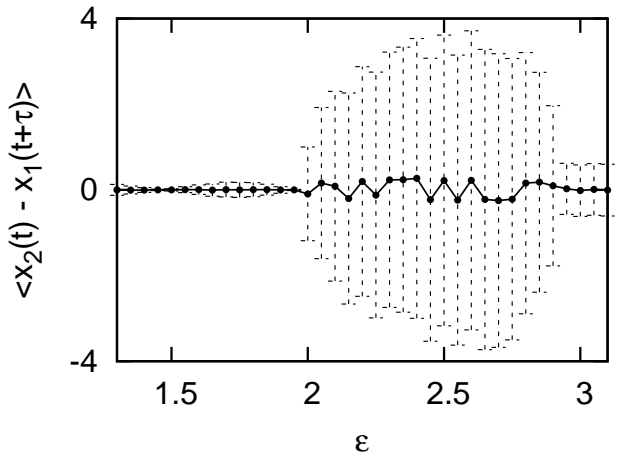}
\caption{\label{error_shift} Average error function computed after shifting one of the time series by half the period in the synchronized periodic regions II - IV of Fig.~\ref{corr_ros}b. The average error $\sim 0$ for $\epsilon < 2$ and $\epsilon > 2.95$ indicating lag synchronization. The region $2 < \epsilon < 2.95$ corresponds to antiphase synchronization in the periodic state.}
\end{figure}
\end{center}

In this section, we present the complete phase diagram in the parameter plane of coupling strengths identifying the regions of different states of synchronization such as complete ( or anti ) synchronization, in-phase ( or antiphase ) synchronization and unsynchronized regions. We use the average phase difference and Lyapunov exponent to mark the different regions of synchronization. In addition, the complete and antisynchronization states are characterized by calculating correltion between the two systems using
\begin{equation}
\label{eq:corr}
C = \frac{ ( x_1(t) - <x_1(t)> ) ( x_2(t) - <x_2(t)> ) }{ \sqrt{ ( x_1(t) - <x_1(t)> )^2 ( x_2(t) - <x_2(t)> )^2 } }
\end{equation}
The phase diagram in the $\epsilon_1$ -- $\epsilon_2$ plane for R\"ossler
system is shown in Fig.~\ref{corr_ros}a for $\beta_1=1, \, \beta_2=-1$. As the coupling strengths increase ( along the diagonal ) we see a transition from the unsynchronized state (dark gray) to the in-phase synchronized state (light gray) and then to the completely synchronized state (white). For large coupling constants, the system becomes unstable (black). The critical coupling constants corresponding to the transitions between the different types of synchronization obey the relation Eq.~(\ref{critical-ep}) as obtained from the stability analysis. Figure.~\ref{corr_ros}b shows a similar phase diagram for $\beta_1=\beta_2=1$. Here dark gray region corresponds to unsynchronized states, region marked I corresponds to antiphase synchronization in chaotic state, regions II - IV corresponds to different regimes of synchronization in periodic states and black region corresponds to unstable states. We find that here, depending on the coupling strength the coupled systems settle to two different periodic states $A$ and $B$. The $x-y$ plane corresponding  to the states $A$ and $B$ are shown in Fig.~\ref{tworos}. In regions II and IV, both are in state $A$ shown in Fig.~\ref{tworos}a. while in region III, one system is in periodic state $A$ and the other in state $B$(Fig.~\ref{tworos}b).
In regions II and IV, the synchronized states are such that $x_1(t+\tau) \simeq x_2(t)$, corresponding to lag synchronization and in region III, the systems are in antiphase synchronization in the periodic state. The average error function calculated after shifting $x_1(t)$ by half the time period for the regions I, III and IV is shown in Fig.~\ref{error_shift}.

\begin{center}
\begin{figure}
\includegraphics[width=0.95\columnwidth]{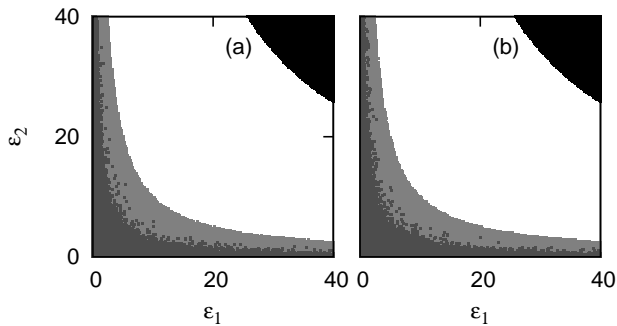}
\caption{\label{corr_lor} Regions of different states of synchronization marked out in parameter plane of coupling strengths $(\epsilon_1, \epsilon_2)$ by computing asymptotic correlation values, average phase difference and similarity functions for Lorenz systems. (a) $\beta_1 = +1,\beta_2 = -1$ (b) $\beta_1 = \beta_2 = +1$. In both cases, white region corresponds to $|C| \sim 0.99 $ indicating synchronized\slash anti-synchronized regions. Light gray region is in-phase ( or antiphase ) syncrhonized state and dark gray region is unsynchronized state. Black region in the upper right corner corresponds to unstable states. In the numerical simulations, $\epsilon_1$ and $\epsilon_2$ are varied in steps of 0.2. The time averages in Eq.~(\ref{eq:corr}) are taken over 50 time units.}
\end{figure}
\end{center}

The similar phase diagrams for coupled Lorenz systems are shown in Figs.~\ref{corr_lor}a and \ref{corr_lor}b. As $\epsilon$ is increased along the diagonal, we observe transitions in the following sequence: unsynchronized state (dark gray) to in-phase\slash antiphase synchronized states (light gray) to complete\slash anti synchronized states (white) to unstable states (black). Here also, the critical coupling constants corresponding to the transitions between the different types of synchronization obey the theoretical relation Eq.~(\ref{critical-ep}).

\section{CONCLUSION}
We report the synchronization of two nonlinear chaotic systems by coupling them indirectly through a common environment. The coupling mechanism proposed is general and can be adjusted for in-phase and antiphase or complete and anti- types of synchronization. The different types of synchronous behavior and the transitions among them are analyzed in the case of two standard systems R\"ossler and Lorenz using the numerically computed Lyapunov exponents,  average phase difference, correlation from time series and similarity function. Using an approximate linear stability analysis, the threshold values of coupling strengths for onset of synchronization of the in-phase or antiphase type are derived and the transitions curves obtained from numerical calculations agree with the curves from stability analysis.

This method of synchronization has the interesting feature that the synchronized state has almost the same phase space structure as that of the uncoupled dynamics. The mehod reported here offers a simple coupling scheme to realize phase ( or antiphase ) synchronization in two coupled chaotic identical systems. As far as we know, the reported works in this are mostly on nonidentical systems with parameter mismatch or delay in coupling. 

The results for synchronized states with such a coupling are presented here for three standard cases such as R\"ossler, Lorenz and Hindmarsh-Rose systems. However, we have checked that it works in general for a few cases also such as FitzHugh Nagumo model of neurons and Mackey-Glass system.


\begin{thebibliography}{1}
\bibitem{Pikovsky2003}
A. S. Pikovsky, M. G. Rosenblum, and J. Kurths, {\em Synchronization: A Universal Concept in Nonlinear Sciences }, Cambridge Nonlinear Science Series, (Cambridge University Press, London, 2003).
\bibitem{Ros96}
M. G. Rosenblum, A.S. Pikovsky, and J. Kurths, Phys. Rev. Lett. {\bf 76}, 1804 (1996). 
\bibitem{Rosa98}
E. R. Rosa, E. Ott, and M. H. Hess, Phys. Rev. Lett. {\bf 80}, 1642 (1998). 
\bibitem{Liu00}
J. Liu, C.Ye, S.Zhang, and W.Song, Phys. Lett. A {\bf 274}, 27 (2000). 
\bibitem{Ros97}
M. G. Rosenblum, A.S. Pikovsky, and J. Kurths, Phys. Rev. Lett. {\bf 78}, 4193 (1997).  
\bibitem{Tahe99}
S. Taherion, and Y. C. Lai, Phys. Rev. E. {\bf 59}, R6247 (1999).  
\bibitem{Vos00}
H. U. Voss, Phys. Rev. E {\bf 61}, 5115 (2000).  
\bibitem{Rul95}
N. F. Rulkov, M. M. Sushchik, L.S. Tsimring, and H.D.I. Abarbanel, Phys. Rev. E {\bf 51}, 980 (1995).
\bibitem{Aba96}
Henry D. I. Abarbanel, Nikolai F. Rulkov, and  M. M. Sushchik, Phys. Rev. E {\bf 53}, 4528 (1996). 
\bibitem{Koc96}
L. Kocarev, and U. Parlitz, Phys. Rev. Lett. {\bf 76}, 1816  (1996).
\bibitem{Fuj83}
H. Fujisaka, and T. Yamada, Prog. Theor. Phys. {\bf 69}, 32 (1983). 
\bibitem{Pec90}
L.M. Pecora, and T. L. Carroll, Phys. Rev. Lett. {\bf 64}, 821 (1990).
\bibitem{Siv01}
S. Sivaprakasam, I. Pierce, P. Rees, P. S. Spencer, K. A. Shore and A. Valle, Phys. Rev. A {\bf 64} ,013805 (2001).
\bibitem{Kim03} 
C.M. Kim, S. Rim, W.H. Kye, J.W. Ryu, and Y.J. Park, Phys. Lett. A, {\bf 320}, 39 (2003).
\bibitem{Zhu07}
H. Zhu, and B. Cui, Chaos {\bf 17}, 043122 (2007).
\bibitem{Ran06} 
A. V. Rangan, and D. Cai, Phys. Rev. Lett. {\bf 96} ,178101 (2006). 
\bibitem{Sin05}
S. Sinha, and S. Sinha, Phys. Rev. E {\bf 71}, 020902(R) (2005).
\bibitem{Che01}
X. Chen and J. E. Cohen, J. Theo. Biol. {\bf 212} , 223 (2001).
\bibitem{Tor01}
R. Toral, C. R. Mirasso, E. Hernandez-Garcia, and O. Piro, Chaos {\bf 11}, 665 (2001). 
\bibitem{Zhou02}
C.S. Zhou, and J. Kurths, Phys. Rev. Lett. {\bf 88} 230602 (2002)
\bibitem{He03}
D. He, P. Shi, and L. Stone, Phys. Rev. E {\bf 67}, 027201 (2003).
\bibitem{Pik97}
A. S. Pikovsky, M. G. Rosenblum, G. V. Osipov, and J. Kurths, Physica D {\bf 104}, 219 (1997).
\bibitem{Pik97chaos}
A.S. Pikovsky, M. Zaks, M. Rosenblum, G. Osipov, and J. Kurths, Chaos {\bf 7}, 680 (1997).
\bibitem{Park99}
E. H. Park, M. A. Zaks, and J. Kurths, Phys Rev E {\bf 60}, 6627  (1999).
\bibitem{Brai94}
Y. Braiman and K. Wiesenfeld, Phys. Rev. B {\bf 49}, 15223 (1994)
\bibitem{Toth06}
R. Toth, A.F. Taylor, and M.R. Tinsley, J. Phys. Chem. B {\bf 110}, 10170 (2006)
\bibitem{Kuz05}
A. Kuznetsov, M. K{\ae}rn and N. Kopell, SIAM J. Appl. Math. {\bf 65}, 392 (2004).
\bibitem{Wang05}
R. Wang, and L. Chen, J. Biol. Rhythms {\bf 20}, 257 (2005).
\bibitem{Gon05}
D. Gonze, S. Bernard, C. Waltermann, A. Kramer and H. Herzel, Biophys. J. {\bf 89}, 120 (2005).
\bibitem{Jav08}
J. Javaloyes, M. Perrin, and A. Politi, Phys. Rev. E {\bf 78}, 011108 (2008).
\bibitem{Kat08}
G. Katriel, Physica D. {\bf 237}, 2933 (2008). 
\bibitem{amb09}
G. Ambika and R. E. Amritkar, Phys. Rev. E {\bf 79}, 056206 (2009).
\bibitem{Wolf85}
 A. Wolf, J. B. Swift, H. L. Swinney, and J. A. Vastano,  Physica D {\bf 16}, 285 (1985).
\bibitem{Boc02}
S. Boccaletti, J. Kurths, G. Osipov, D. L. Valladares, and C. S. Zhou
Phys. Rep. {\bf 366}, 1 (2002). 
\bibitem{Sen09}
D. V. Senthilkumar, J. Kurths, and M. Lakshmanan, Chaos {\bf 19}, 023107 (2009).
\end{thebibliography}
\end{document}